\documentclass[a4paper,11pt]{article}
\usepackage[T1]{fontenc}
\usepackage[a4paper,textwidth=440pt,textheight=666pt,centering,
            headheight=13pt,footskip=30pt]{geometry}
\usepackage{graphicx}
\DeclareGraphicsExtensions{.pdf}
\usepackage{amsmath,amssymb}
\usepackage{booktabs}
\usepackage{placeins}
\usepackage[numbers,sort&compress]{natbib}
\usepackage[font=small,labelfont=bf,labelsep=period]{caption}
\usepackage[colorlinks=true,allcolors=blue,linktocpage=true]{hyperref}
\numberwithin{equation}{section}
\newenvironment{acknowledgments}{\section*{Acknowledgements}}{}
\newcommand{\dd}{\mathrm{d}}
\newcommand{\rs}{r_*}
\newcommand{\Wodd}{\mathcal W_{\rm o}}
\newcommand{\Weven}{\mathcal W_{\rm e}}
\newcommand{\aM}{\widehat\alpha}
\newcommand{\wM}{\widehat\omega}


\renewenvironment{thebibliography}[1]{%
  \begin{originalthebibliography}{#1}\small\raggedright
  \setlength{\itemsep}{5pt plus 0.2ex minus 0.05ex}%
}{\end{originalthebibliography}}
\hypersetup{pdftitle={Charge-Activated Birefringence and Shadow Mimicry around Reissner-Nordstr\"om Black Holes},pdfauthor={Mao-Yuan Wan and Chen Wu},pdfkeywords={GR black holes, Quantum fields in curved spacetimes, gravitational lensing, absorption and radiation processes}}

\begin{document}
\begin{titlepage}
\thispagestyle{empty}
\vspace*{8mm}
\begin{center}
{\LARGE\bfseries Charge-Activated Birefringence and Shadow Mimicry around Reissner-Nordstr\"om Black Holes\par}
\vspace{8mm}
{\large Mao-Yuan Wan\qquad Chen Wu$^{\ast}$\par}
\vspace{4mm}
{\small\itshape Xingzhi College, Zhejiang Normal University, Jinhua 321004, Zhejiang, China\par}
\vspace{4mm}
{\small E-mail: \href{mailto:maoyuan.wan.physics@gmail.com}{maoyuan.wan.physics@gmail.com}\par}
{\small E-mail: \href{mailto:wuchenoffd@gmail.com}{wuchenoffd@gmail.com}\par}
\smallskip
{\small $^{\ast}$Corresponding author.\par}
\end{center}
\vspace{7mm}
\noindent\textbf{Abstract}\par\smallskip
\noindent
We demonstrate that electric charge can activate birefringence for a neutral vector field coupled to the Einstein tensor around a Reissner-Nordstr\"om black hole.
Changing the electric charge of an uncoupled black hole can mimic one parity's shadow radius of a coupled black hole.
The coupling makes the two parity looks differently.
It separates their shadow edges of two parities and produces a backward-scattered signal with reversed helicity.
However changing the charge alone cannot reproduce both effects.
They require both nonzero charge and coupling.
At weak coupling,
the leading correction is the same for both potentials.
Positive coupling raises the potential,
reduces low-frequency absorption and decreases the average shadow radius.
Negative coupling reverses these trends.
The backward signal is nonzero, this is due to the difference between the complex scattering amplitudes of the two parities.
It is entirely helicity reversed.
For the uncoupled field,
the shadows are the same for both parities and the backward intensity vanishes exactly at every charge.
We can use the split edge and backward signal to distinguish the coupling from a change of charge alone.
A comparison with observations still requires a mechanism that produces and detects this field.

\end{titlepage}
\pagenumbering{arabic}
\tableofcontents
\medskip

\section{Introduction}\label{sec:1}

A change in shadow radius may originate from either spacetime geometry or field propagation.
By comparing absorption, scattering and parity-dependent capture radii, we distinguish curvature coupling from changes in the black hole charge.

Curvature couplings change how vector fields propagate and can make absorption,
scattering and lensing depend on polarisation~\cite{ref01,ref02}.
The effective potentials of Weyl-coupled photons on vacuum black-hole backgrounds were obtained in earlier work~\cite{ref03}.
This coupling was related to lensing via weak deflection and strong deflection~\cite{ref04,ref05}.
Earlier studies examined double shadows~\cite{ref06},
polarised thin-disc images~\cite{ref07},
and Kerr shadows cast by extraordinary light rays~\cite{ref08}.
Reference~\cite{ref09} calculated absorption, scattering and shadows for Weyl-coupled photons.
It found that the two parities have different shadow edges and the coupling produces a helicity-reversing signal in the backward direction.
These effects occur without charge because the Schwarzschild Weyl tensor is nonzero.
Here we consider the interaction $\alpha G^{\mu\rho}F_{\mu\nu}F_\rho{}^\nu$,
which couples a neutral vector test field to the Einstein tensor.
$\alpha$ is the coupling constant,
$G^{\mu\rho}$ is the Einstein tensor and $F_{\mu\nu}$ is the test-field strength.
For a Reissner-Nordstr\"om (RN) black hole,
the supporting electromagnetic field makes the Einstein tensor nonzero and activates the propagation correction.
The interaction vanishes at zero charge.

Riemann-tensor couplings yield polarisation dependent photon rings~\cite{ref10}.
Vacuum polarisation on a deformed compact-object background changes photon velocities and lensing~\cite{ref11}.
Hu et al. showed that QED corrections change shadow sizes and shapes for static black holes with magnetic monopoles or surrounding magnetic fields~\cite{ref12}.
Zhong et al. extended this analysis to Kerr black holes in uniform magnetic fields using backward ray tracing~\cite{ref13}.

Nonlinear electrodynamics provides another mechanism for polarisation-dependent propagation.
Murk and Soranidis showed that birefringence modifies the light-ring structure of regular black holes and horizonless ultracompact objects through distinct optical metrics~\cite{ref14}.
De Paula et al. obtained two polarisation-dependent shadow radii for magnetically charged black holes in an Euler--Heisenberg model~\cite{ref15}.
We keep the RN geometry fixed.
The interaction modifies only test-field propagation.

Chen and Jing studied this vector model and its perturbation equations~\cite{ref16}.
Liao,
Wei and Chen studied parity-dependent absorption and Hawking radiation in this model~\cite{ref17}.
We integrate the radial equations directly, from low frequencies through the transmission transition.
We also retain the complex reflection amplitudes needed for differential scattering.

The partial-wave description was established by early calculations of scalar scattering and absorption by Schwarzschild black holes~\cite{ref18,ref19}.
The role of scattering phases was clarified by phase integral methods and studies of rotating backgrounds~\cite {ref20,ref21}.
On the RN background, scalar fields coupled to the Einstein tensor possess modified greybody factors~\cite {ref22}.
The vector field separates into two sets of spherical modes distinguished by their behaviour under spatial inversion.
These sets are called odd and even parity.
The birefringence studied here arises from their different propagation.
Comparing them gives polarisation information that a scalar channel does not have.

Charged fields can exhibit superradiance even on a static background.
Dolan et al. found superradiant quasibound states of a charged massive scalar around an electrically charged Ay\'on-Beato--Garc\'ia regular black hole~\cite{ref23}.
Our neutral field has no such amplification.

We treat the vector field as distinct from the Maxwell field that supports the RN geometry.
This choice matters for backward scattering.
Electromagnetic perturbations coupled to gravitational perturbations of a charged black hole can already produce helicity reversal without the interaction studied here~\cite{ref24}.
In our neutral test-field equations,
the uncoupled parity potentials coincide and the reversing amplitude vanishes.
The spin-one backward null on Schwarzschild~\cite{ref25} therefore carries over to every charge in our uncoupled reference model.

The shadow imaging proposal of Ref.~\cite{ref26} and the Event Horizon Telescope images of M87* and Sgr~A*~\cite{ref27,ref28} motivate comparisons with photon ring.
The study of polarisation dependence is also motivated by the resolved ring polarisation of M87*~\cite {ref29}.
These observations probe ordinary electromagnetic radiation.
However, our study describes the capture of a separate test field.

Greybody factors give the fraction of each incoming mode absorbed by the black hole.
Scattering also depends on reflection phases and can reverse the handedness of circular polarisation,
an effect called helicity reversal.
In geometric optics,
the two critical impact parameters define the parity capture edges.
We first compare coupled and uncoupled fields at fixed mass and charge,
then vary the uncoupled RN charge to test shadow-radius matching.
The geometric interpretation of these results is provided by work on critical curves and lensing rings~\cite{ref30} and reviews of shadows and strong lensing~\cite{ref31,ref32}.

Section~\ref{sec:2} gives the model, potentials and optical quantities.
Section~\ref{sec:3} defines the observables and the partial-wave calculation.
Sections~\ref{sec:4} to~\ref{sec:6} present absorption,
scattering and shadow geometry.
We discuss charge matching and the physical limitations in Section~\ref{sec:7}.
Section~\ref{sec:8} gives our conclusions.

\section{Model and parity potentials}\label{sec:2}

\subsection{Background and radial equations}

We use geometrised units $G=c=1$,
signature $(+,-,-,-)$,
and $R^\rho{}_{\sigma\mu\nu}=\partial_\mu\Gamma^\rho{}_{\nu\sigma} -\partial_\nu\Gamma^\rho{}_{\mu\sigma} +\Gamma^\rho{}_{\mu\lambda}\Gamma^\lambda{}_{\nu\sigma} -\Gamma^\rho{}_{\nu\lambda}\Gamma^\lambda{}_{\mu\sigma}$,
with $R_{\sigma\nu}=R^\rho{}_{\sigma\rho\nu}$.
The symbols $\Gamma^\rho{}_{\mu\nu}$ are the Christoffel symbols of the background metric $g_{\mu\nu}$.
The coupling $\alpha$ has dimensions of length squared.
The action combines the Maxwell kinetic term with an Einstein-tensor interaction:
\begin{equation}
 S=\int\dd^4x\sqrt{-g}\left[-\frac14 F_{\mu\nu}F^{\mu\nu}
 +\alpha G^{\mu\rho}F_{\mu\nu}F_\rho{}^\nu\right],
 \qquad F_{\mu\nu}=\partial_\mu A_\nu-\partial_\nu A_\mu .
 \label{eq:2.1}
\end{equation}
Here $A_\mu$ denotes the vector potential and $g=\det g_{\mu\nu}$.
Varying the action yields $\nabla_\mu P^{\mu\nu}=0$,
where $\nabla_\mu$ is the metric-compatible covariant derivative and the constitutive tensor is
\begin{equation}
 P^{\mu\nu}=F^{\mu\nu}-2\alpha G^{\mu\rho}F_\rho{}^\nu
 +2\alpha G^{\nu\rho}F_\rho{}^\mu .
 \label{eq:2.2}
\end{equation}
We keep the background metric fixed:
\begin{equation}
 \dd s^2=f\,\dd t^2-f^{-1}\dd r^2-r^2\dd\Omega^2,\qquad
 f=1-\frac{2M}{r}+\frac{Q^2}{r^2},\qquad
 r_\pm=M\pm\sqrt{M^2-Q^2}.
 \label{eq:2.3}
\end{equation}
The coordinate $r$ is the areal radius,
and $\dd\Omega^2=\dd\theta^2+\sin^2\theta\,\dd\phi^2$ is the line element on the unit sphere.
The radius $r_+$ is the outer horizon.
For $Q\ne0$,
$r_-$ is the inner horizon.
We denote the positive frequency measured in the asymptotic time $t$ by $\omega$.
We write $q=Q/M$,
$\aM=\alpha/M^2$ and $\wM=M\omega$.
These dimensionless parameters are used in all figures; a figure label $\alpha$ denotes its numerical value in units of $M^2$.

We treat $A_\mu$ and the Maxwell field charge $Q$ as separate fields.
The stress-energy tensor of $A_\mu$ is neglected in the background equations.
Changing $\alpha$ therefore changes propagation while keeping the metric and both horizon radii fixed.
If the two fields were identified, one would need a separate calculation that accounts for backreaction.
At $Q=0$,
the Einstein tensor vanishes identically.
The Schwarzschild limit thus holds for all finite $\alpha$,
without a weak-coupling expansion.

We decompose the field into spherical harmonics~\cite{ref33} with time dependence $e^{-i\omega t}$.
We label odd and even parity by $p={\rm o,e}$.
Each multipole $\ell$ describes an angular pattern on the sphere.
The field equation is reduced to radial scattering problems once these patterns are separated.
For each multipole $\ell\geq1$,
the radial master amplitude $\Psi_\ell^p(r)$ satisfies
\begin{equation}
 \frac{\dd^2\Psi_\ell^p}{\dd\rs^2}
 +\left[\omega^2-V_\ell^p(r)\right]\Psi_\ell^p=0,\qquad
 \frac{\dd\rs}{\dd r}=\frac1f,
 \label{eq:2.4}
\end{equation}
where $\rs$ is the tortoise coordinate.
It sets the outer horizon at $\rs\to-\infty$,
so the horizon boundary condition takes the form of a simple travelling wave.
The potential forms a barrier between an incoming wave and the horizon.
Its height and shape determine how much of the wave is transmitted or reflected.
The two parity potentials are
\begin{equation}
 V_\ell^p=\frac{f\ell(\ell+1)}{r^2}\mathcal W_p,\qquad
 \Wodd=1+x,\qquad \Weven=\frac1{1-x},\qquad
 x=\frac{4\alpha Q^2}{r^4}.
 \label{eq:2.5}
\end{equation}

The denominator of the even-parity differs in sign from the printed Eq.~(16) of Ref.~\cite{ref16}.
On RN,
$G^t{}_t=G^r{}_r=Q^2/r^4$ and $G^\theta{}_\theta=G^\phi{}_\phi=-Q^2/r^4$.
Equation~\eqref{eq:2.2} therefore gives $P^{tr}=-(1-x)F_{tr}$ and $P^{\theta\phi}=(1+x)F^{\theta\phi}$; the mixed angular-radial and angular-time components are unchanged.
For an even mode,
we write $A_t=jY e^{-i\omega t}$,
$A_r=hY e^{-i\omega t}$ and $A_a=k\,\partial_aY e^{-i\omega t}$,
with $a=\theta,\phi$.
Here $Y=Y_{\ell m}(\theta,\phi)$ is a scalar spherical harmonic,
$m$ is its azimuthal index and $j(r)$,
$h(r)$ and $k(r)$ are radial coefficient functions.
We define $\Phi=j+i\omega k$,
$Z=h-k'$,
$E=-(\Phi'+i\omega Z)$ and $\chi=r^2(1-x)E$,
where primes denote $\dd/\dd r$.
The amplitude $E$ gives $F_{tr}=EYe^{-i\omega t}$,
and we may choose $\Psi_\ell^{\rm e}=\chi$.
The radial constraints reduce to
\begin{equation}
 Z=\frac{i\omega\chi}{\ell(\ell+1)f},\qquad
 \Phi=-\frac{f\chi'}{\ell(\ell+1)}.
 \label{eq:2.6}
\end{equation}
Substitution into $E=-(\Phi'+i\omega Z)$ gives
\begin{equation}
 \frac{\dd^2\chi}{\dd\rs^2}
 +\left[\omega^2-\frac{f\ell(\ell+1)}{r^2(1-x)}\right]\chi=0 .
 \label{eq:2.7}
\end{equation}
The odd angular equation carries the factor $1+x$ in its centrifugal term.
Thus the same constitutive equation fixes both factors in Eq.~\eqref{eq:2.5}.

\subsection{Barrier changes and parameter choices}

The relation
\begin{equation}
 \Weven-\Wodd=\frac{x^2}{1-x},\qquad
 \Wodd=1+x,\qquad \Weven=1+x+x^2+O(x^3)
 \label{eq:2.8}
\end{equation}
shows that the common barrier shift is first order in $x$,
while parity splitting begins at second order.
Positive coupling raises both barriers and negative coupling lowers them.
For $1-x>0$,
the even barrier is always higher.
This explains the main transmission trends;
angular interference also depends on reflection phases.

\begin{figure}[!htbp]
 \centering\includegraphics[width=\textwidth]{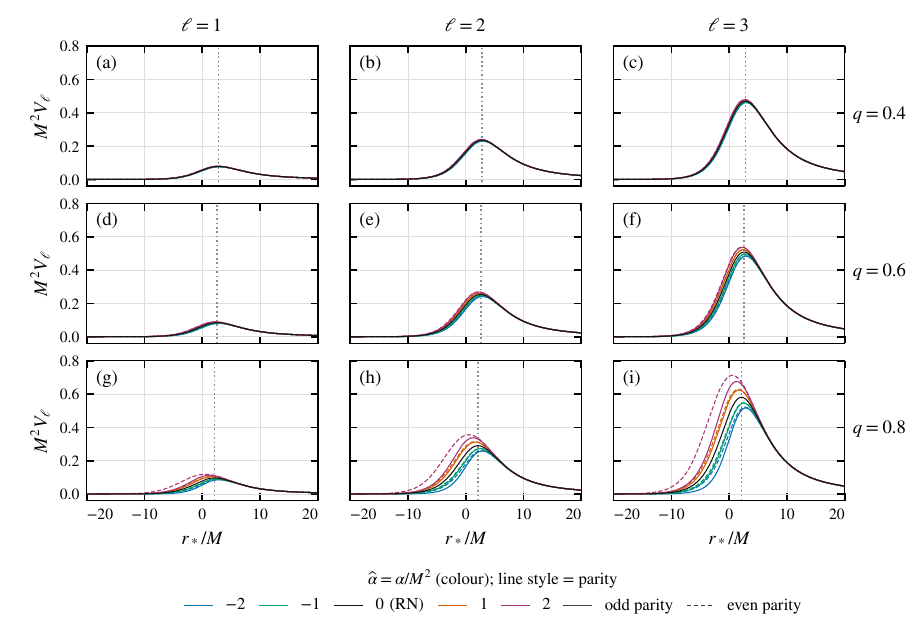}
 \caption{Effective potentials versus tortoise coordinate:
rows $q=0.4,0.6,0.8$ and columns $\ell=1,2,3$.
Blue,
green,
black,
orange and purple denote $\aM=-2,-1,0,1,2$.
Solid/dashed curves show odd/even parity;
black is the common RN control.
All panels share a vertical scale.
Dotted lines mark the same-charge RN peak using Eq.~\eqref{eq:2.10}.}
 \label{fig:1}
\end{figure}

We use RN black holes with $|Q|<M$.
We also require both parity equations to remain regular outside the outer horizon:
\begin{equation}
 |\alpha|<\frac{r_+^4}{4Q^2}\quad (Q\ne0).
 \label{eq:2.9}
\end{equation}
This condition keeps both optical factors positive outside the horizon.
Figures~\ref{fig:1} to~\ref{fig:7} use $q=0.4,0.6,0.8$ with $\aM=-2,-1,0,1,2$,
giving fifteen parameter pairs.
Tables~\ref{tab:1} to~\ref{tab:3} add the intermediate half-integer couplings,
giving twenty-seven pairs.
All these pairs satisfy Eq.~\eqref{eq:2.9}.

The local coupling magnitude at the horizon is $\epsilon_h=4|\alpha|Q^2/r_+^4$.
Its largest value on this grid is $0.78125$,
at $q=0.8$ and $|\aM|=2$.
This correction is not much smaller than unity.
We therefore treat the larger-coupling cases as finite-coupling results of the specified model.
Additional interactions might be required by a generic effective field theory at such coupling strengths.

Figure~\ref{fig:1} compares the three lowest propagating multipoles.
Barrier height scales with $\ell(\ell+1)$,
but the peak location is independent of $\ell$.
Increasing charge enhances the common shift and parity separation.
We use the following convention for both the potential plots and the wave calculation by
\begin{equation}
 \frac{\rs}{M}=R+\frac{R_+^2}{R_+-R_-}\ln(R-R_+)
 -\frac{R_-^2}{R_+-R_-}\ln(R-R_-),
 \quad R=\frac rM,\quad R_\pm=\frac{r_\pm}{M}.
 \label{eq:2.10}
\end{equation}

\subsection{Ray capture and deflection}

At short wavelengths,
we can describe the field by light rays.
Their trajectories are governed by the optical metric, which can differ for the two parities.
It describes propagation on the fixed RN background; it does not replace the background in the gravitational equations.
We can obtain it from the large-$\ell$ limit of Eq.~\eqref{eq:2.4},
up to a conformal factor:
\begin{equation}
 \dd s_p^2=f\,\dd t^2-f^{-1}\dd r^2
 -\frac{r^2}{\mathcal W_p}\dd\Omega^2 .
 \label{eq:2.11}
\end{equation}
We label an incoming ray by its impact parameter,
its transverse offset measured far from the black hole.
The critical value $b_c^p$ separates capture from escape and sets the shadow radius.
At this critical value,
a ray approaches an unstable circular orbit of radius $r_{\rm ph}^p$.
With $H_p=f\mathcal W_p/r^2$,
the two quantities satisfy
\begin{equation}
 H_p'(r_{\rm ph}^p)=0,\qquad
 b_c^p=\left[H_p(r_{\rm ph}^p)\right]^{-1/2}.
 \label{eq:2.12}
\end{equation}
We select the exterior maximum of $H_p$,
which sets the capture area $\pi(b_c^p)^2$.
Its height and curvature govern the wave-transmission threshold and width.
At $Q=0$,
both branches give $r_{\rm ph}=3M$ and $b_c=3\sqrt3M$ for all finite $\alpha$.
These critical rays underlie classical photon-escape analysis~\cite{ref34}.
The glory impact parameter $b_g^p$ instead selects backward-deflected rays,
$\Theta_p(b_g^p)=\pi$,
and controls near-backward scattering.

\FloatBarrier

\section{Scattering observables}\label{sec:3}

A unit-amplitude ingoing wave is imposed at the horizon.
At infinity,
the solution is matched to incoming and outgoing modes $u_{\rm in}$ and $u_{\rm out}$ with complex amplitudes $A_{\rm in}^p$ and $A_{\rm out}^p$:
\begin{equation}
 \Psi_\ell^p\sim e^{-i\omega\rs}\quad(\rs\to-\infty),\qquad
 \Psi_\ell^p\sim A_{\rm in}^p u_{\rm in}
 +A_{\rm out}^p u_{\rm out}\quad(\rs\to+\infty).
 \label{eq:3.1}
\end{equation}
We suppress the multipole index in these amplitudes.
The asymptotic basis includes the centrifugal phase,
$u_{\rm out}\sim\exp[i\omega\rs-i(\ell+1)\pi/2]$ and $u_{\rm in}=u_{\rm out}^*$.
Under this convention
\begin{equation}
 S_\ell^p=\frac{A_{\rm out}^p}{A_{\rm in}^p},\qquad
 \Gamma_\ell^p=\frac1{|A_{\rm in}^p|^2}=1-|S_\ell^p|^2.
 \label{eq:3.2}
\end{equation}
The reflected amplitude, together with its phase, is encoded in the complex quantity $S_\ell^p$ relative to the incident amplitude.
It is the partial-wave scattering matrix.
The greybody factor $\Gamma_\ell^p$ is the transmitted fraction of incident flux.
We evaluate it from the incoming amplitude.

The horizon and incident fluxes share the same positive frequency $\omega$.
Our static background and neutral field produce no horizon frequency shift,
so there is no superradiant amplification.
Equation~\eqref{eq:3.2} gives a transmission probability between zero and one.
The neutral field carries no electromagnetic Coulomb interaction.
We nevertheless retain the long-range gravitational phase in the asymptotic matching and angular sum.

We obtain the absorption cross section by summing the transmitted flux over all angular modes~\cite{ref35}.
The factor $2\ell+1$ accounts for the angular states of each multipole.
For the unpolarised result,
we average the two parity cross sections:
\begin{equation}
 \sigma_{\rm abs}^p=\frac{\pi}{\omega^2}
 \sum_{\ell=1}^\infty(2\ell+1)\Gamma_\ell^p,\qquad
 \sigma_{\rm abs}^{\rm unpol}=\frac{\sigma_{\rm abs}^{\rm o}+\sigma_{\rm abs}^{\rm e}}2.
 \label{eq:3.3}
\end{equation}
We quantify the relative parity difference by
\begin{equation}
 \mathcal A_{\rm abs}=
 \frac{\sigma_{\rm abs}^{\rm o}-\sigma_{\rm abs}^{\rm e}}
 {\sigma_{\rm abs}^{\rm o}+\sigma_{\rm abs}^{\rm e}}
 \label{eq:3.4}
\end{equation}
A positive value indicates that odd parity absorbs more than even parity.
A zero value means that their total absorbed fluxes are equal.

To find how much radiation leaves in each direction,
we must also keep track of the relative phases of the reflected modes.
Their waves add or cancel at different angles.
This is why differential scattering requires the complex $S$ matrices rather than transmission probabilities alone.
We use $\theta$ for the angle between the incident and outgoing directions,
with $\theta=0$ corresponding to forward scattering and $\theta=\pi$ backward scattering.

Let $z=\cos\theta$,
$\pi_\ell=P_\ell'(z)$ and $\tau_\ell=zP_\ell'(z)-(1-z^2)P_\ell''(z)$,
where $P_\ell$ is the Legendre polynomial and the primes indicate differentiation with respect to $z$.
Helicity is the handedness of circular polarisation relative to propagation.
The spin-1 helicity-preserving ($+$) and helicity-reversing ($-$) amplitudes are
\begin{align}
 F_\pm(\theta)&=\sum_{\ell=1}^\infty
 \frac{2\ell+1}{\ell(\ell+1)}g_\ell^\pm
 \left(\pi_\ell\pm\tau_\ell\right),\label{eq:3.5}\\
 g_\ell^+&=\frac{S_\ell^{\rm e}+S_\ell^{\rm o}}2-1,\qquad
 g_\ell^-=\frac{S_\ell^{\rm o}-S_\ell^{\rm e}}2 .\nonumber
\end{align}
An overall phase convention does not affect the intensities.
The helicity-preserving and helicity-reversing contributions give
\begin{equation}
 I_{\rm P}=\frac{|F_+|^2}{4\omega^2},\qquad
 I_{\rm R}=\frac{|F_-|^2}{4\omega^2},\qquad
 I\equiv\frac{\dd\sigma}{\dd\Omega}=I_{\rm P}+I_{\rm R},\qquad
 \eta_{\rm R}=\frac{I_{\rm R}}I .
 \label{eq:3.6}
\end{equation}
Here $\dd\Omega$ is the solid-angle element of the scattered direction.
Thus $I$ is the total differential cross section,
and $\eta_{\rm R}$ gives the fraction carried by the helicity-reversing contribution.
This fraction is defined only for $I>0$.
At exact backward scattering,
$\pi_\ell+\tau_\ell=0$ for each multipole,
so $I(\pi)=I_{\rm R}(\pi)$.
The exact backward direction therefore isolates helicity reversal.
When $\alpha=0$,
$S_\ell^{\rm o}=S_\ell^{\rm e}$ and the backward intensity is exactly zero at every charge in this test-field model.

We obtain the partial-wave amplitudes by direct Numerov integration of the full radial equation~\eqref{eq:2.4}.
For the greybody factors and differential scattering at $\ell=1,2,3$,
the exterior solution is matched to an RN asymptotic Jost basis.
The total absorption calculation uses Riccati-Hankel matching together with a residual-tail Born phase correction.
We evaluate the angular sums with series reduction~\cite{ref36,ref37}.

\section{Greybody factors and absorption}\label{sec:4}

\subsection{Barrier transmission for \texorpdfstring{$\ell=1,2,3$}{ell = 1, 2, 3}}

\begin{figure}[!htbp]
 \centering\includegraphics[width=\textwidth]{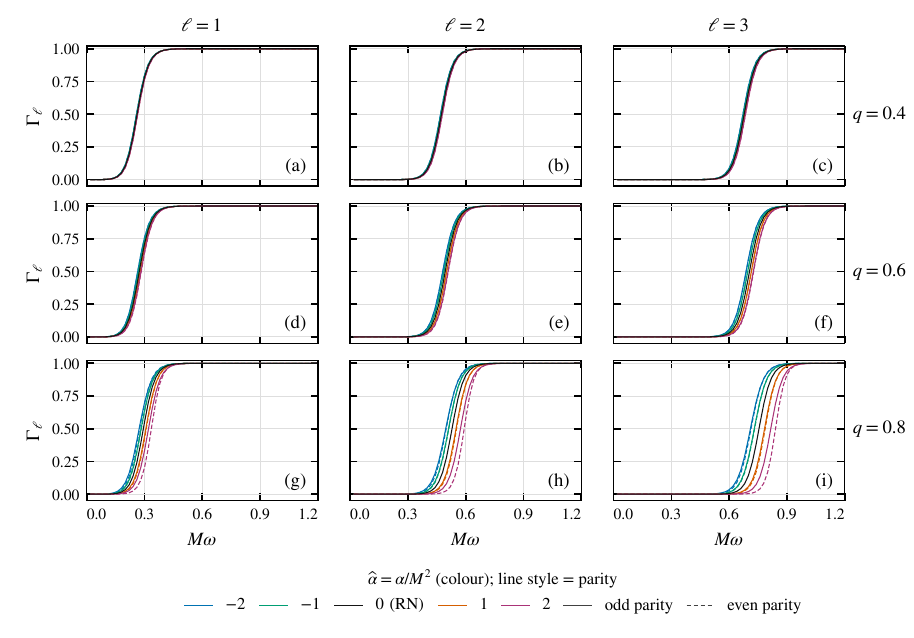}
 \caption{Greybody factors with the layout and coupling colours of Figure~\ref{fig:1}.
Solid/dashed curves show odd/even parity;
the black RN curve represents both.
All panels share linear scales.}
 \label{fig:2}
\end{figure}

Figure~\ref{fig:2} shows the transition from reflection to nearly unit transmission.
Increasing $\ell$ or choosing positive coupling raises the barrier and shifts this transition to higher frequency.
Negative coupling shifts it lower.
The higher even-parity barrier gives lower transmission in the transition region.
The common shift dominates at moderate coupling;
the separation grows at larger charge and is greatest for $\aM=2$ on this grid.
At $\aM=0$,
both parity curves coincide.

\begin{figure}[!htbp]
 \centering\includegraphics[width=\textwidth]{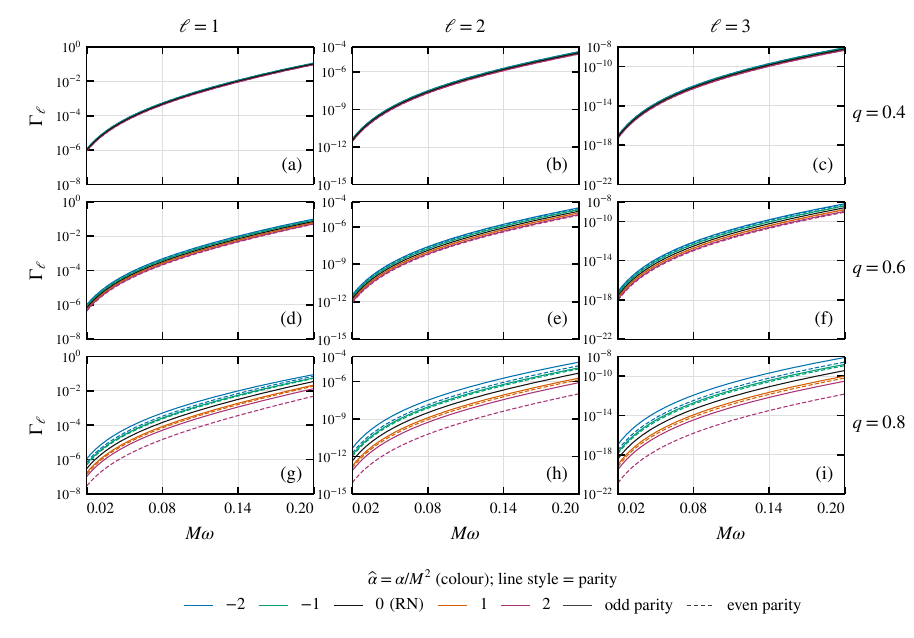}
 \caption{Low-frequency greybody factors over $0.02\leq\wM\leq0.20$,
with the layout,
colours and line styles of Figure~\ref{fig:1}.
Logarithmic vertical limits are shared within each multipole column.
Curves connect numerical samples rather than asymptotic power-law fits.}
 \label{fig:3}
\end{figure}

Figure~\ref{fig:3} resolves the low-frequency tails on logarithmic axes.
Higher multipoles are suppressed,
so $\ell=1$ dominates the absorption.
The minimally coupled scalar horizon-area limit~\cite{ref38} does not apply to this vector sum.
The plotted interval is finite and does not determine an exact asymptotic power law.

Low-frequency transmission depends on tunnelling across the forbidden radial interval,
whereas $b_c$ depends only on the maximum of $H_p$.
A small barrier difference can therefore produce a large relative transmission contrast,
even when both fluxes are weak.
This explains why absorption asymmetry can exceed the fractional shadow separation.

\subsection{Unpolarised absorption and parity asymmetry}

\begin{figure}[!htbp]
 \centering\includegraphics[width=\textwidth]{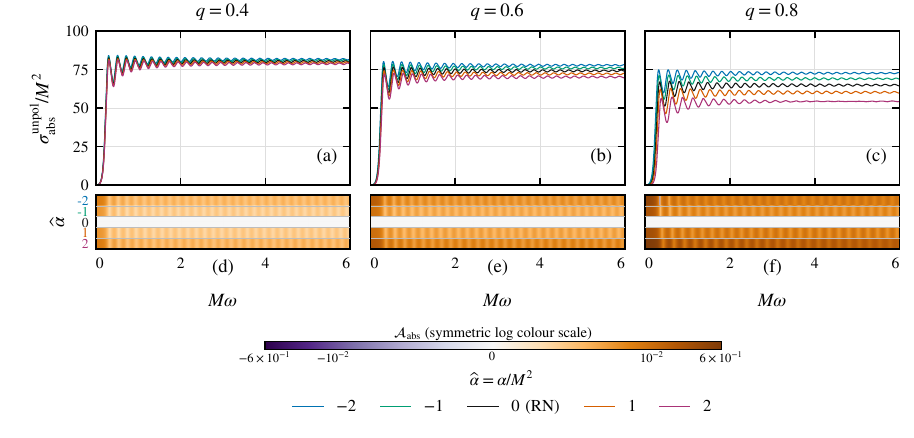}
 \caption{Unpolarised absorption (top) and parity asymmetry (bottom) for $q=0.4,0.6,0.8$ (columns).
Solid-curve colours follow Figure~\ref{fig:1}.
The total absorption includes higher multipoles.
Strips run from $\aM=-2$ to $2$ in unit steps,
top to bottom.
Their shared symmetric logarithmic scale is linear for $|\mathcal A_{\rm abs}|\leq10^{-5}$;
purple/orange-brown denote negative/positive values.
RN asymmetry is zero.}
 \label{fig:4}
\end{figure}

Figure~\ref{fig:4} compares total absorption and parity asymmetry.
At $q=0.8$ and $\wM=1$,
$\aM=1$ reduces absorption by $13.14\%$ relative to RN,
whereas $\aM=-1$ increases it by $7.27\%$.
The corresponding changes at $\aM=2$ and $-2$ are $-25.20\%$ and $+11.04\%$ (Table~\ref{tab:2}).
This sign dependence reflects the nonlinear even-parity factor and transmission probability.

At the same charge and frequency,
$\mathcal A_{\rm abs}=0.081\%,0.837\%,2.881\%$ for $\aM=-1,1,2$.
The low-frequency contrast can be much larger:
at $\wM=0.02$ and $\aM=2$,
it reaches $56.2\%$,
but the unpolarised cross section is only $1.60\times10^{-3}M^2$.
At $\aM=1$ and $-1$,
the asymmetries at this frequency are $9.60\%$ and $6.52\%$.
They fall to $2.269\%$ and $0.428\%$ at $\wM=0.30$.
Thus a large relative contrast need not imply a large absorbed flux.

The response strengthens with charge.
For $\aM=2$ and $\wM=1$,
the absorption reduction grows from $1.10\%$ at $q=0.4$ to $25.20\%$ at $q=0.8$;
the asymmetry grows from $0.0119\%$ to $2.8813\%$.
Although its leading weak-coupling term is even in $\alpha$,
the full asymmetry need not be.
The $q=0.8$,
$\aM=-2$ strip also has a narrow sign reversal:
$\mathcal A_{\rm abs}\simeq-0.0483\%$ at $\wM=0.338$.
Here even parity absorbs slightly more despite its higher potential.
Barrier-height ordering alone therefore does not fix total absorption ordering at every frequency.

At high frequency,
the unpolarised cross section oscillates around the mean parity capture area,
\begin{equation}
 \sigma_{\rm geo}^{\rm unpol}
 =\frac{\pi}{2}\left[(b_c^{\rm o})^2+(b_c^{\rm e})^2\right].
 \label{eq:4.1}
\end{equation}
Oscillations around this classical scale arise from wave interference,
as found in Schwarzschild absorption~\cite{ref19} and developed in the analysis of high-energy fine structure~\cite{ref39}.
For the fifteen parameter pairs in Figure~\ref{fig:4},
the absorption averaged over $4\leq\wM\leq6$ agrees with the geometric value to within $0.122\%$.
Oscillations around this average persist at individual frequencies.
The high-frequency comparison links the absorption spectrum to the capture radii discussed in Section~\ref{sec:6}.

\subsection{high-frequency absorption oscillates}
\label{sec:4.3}

High-frequency oscillations are associated with rays near the unstable circular orbit.
We denote its orbital frequency by $\Omega_p$ and the exponential growth rate of radial ray displacements by $\Lambda_p$.
Both use asymptotic time $t$:
\begin{equation}
 \Omega_p=\frac1{b_c^p},\qquad
 \Lambda_p=f(r_{\rm ph}^p)
 \sqrt{-\frac{H_p''(r_{\rm ph}^p)}{2H_p(r_{\rm ph}^p)}},\qquad
 \eta_p=\frac{\Lambda_p}{\Omega_p}.
 \label{eq:4.2}
\end{equation}
Primes denote radial derivatives.
The ratio $\eta_p$ compares radial escape with orbital motion;
it does not describe a growing field perturbation.

At short wavelengths,
these orbital rates give the leading oscillation of each parity cross section.
Using the photon sphere approximation from Ref.~\cite{ref39}, we get
\begin{equation}
 \frac{\sigma_{\rm abs}^p}{\pi(b_c^p)^2}-1
 \simeq -8\pi\eta_p e^{-\pi\eta_p}
 \operatorname{sinc}(2\pi\omega b_c^p),\qquad
 \operatorname{sinc}z=\frac{\sin z}{z}.
 \label{eq:4.3}
\end{equation}
This short wavelength, or eikonal, approximation is used to interpret the high-frequency partial-wave results.
We compute the greybody factors from the full radial equations.
The sinc gives the phase accumulated over a circular orbit.
Its coefficient depends on the escape rate from that orbit.
Both vector potentials have the leading form $\ell(\ell+1)H_p$,
so this construction works for both branches at leading eikonal order.

Two models can have the same capture radius and hence the same leading oscillation period.
However, their oscillation amplitudes can differ, since rays leave the circular orbit at different rates.
The leading period is $\Delta\omega=1/b_c^p=\Omega_p$.
The coefficient is $8\pi\eta_p e^{-\pi\eta_p}$.
At $(q,\aM)=(0.8,1)$,
we obtain $\eta_{\rm o}=0.8350$ and $\eta_{\rm e}=0.8095$.
At $(0.8,2)$,
the values are $0.8035$ and $0.6501$.
They give leading oscillation coefficients of $1.618$ and $2.119$.
So the even branch can have a smaller average capture area while its relative oscillation coefficient is larger.
These coefficients multiply the sinc function.
By themselves, they do not represent fractional changes of the cross section.

The geometric limit of the absorption asymmetry is
\begin{equation}
 \mathcal A_{\rm geo}=
 \frac{(b_c^{\rm o})^2-(b_c^{\rm e})^2}
 {(b_c^{\rm o})^2+(b_c^{\rm e})^2}.
 \label{eq:4.4}
\end{equation}
For $q=0.8$,
this is $0.156\%$,
$0.354\%$ and $2.621\%$ at $\aM=-1,1,2$.
Interference oscillations are also present in the finite-frequency values shown in Figure~\ref{fig:4},
so their approach to this limit need not be monotonic.
These geometric quantities are used to interpret the spectral scales.
\FloatBarrier

\section{Differential scattering and backward intensity}\label{sec:5}

\subsection{Scattering angles and helicity reversal}

\begin{figure}[!htbp]
 \centering\includegraphics[width=\textwidth]{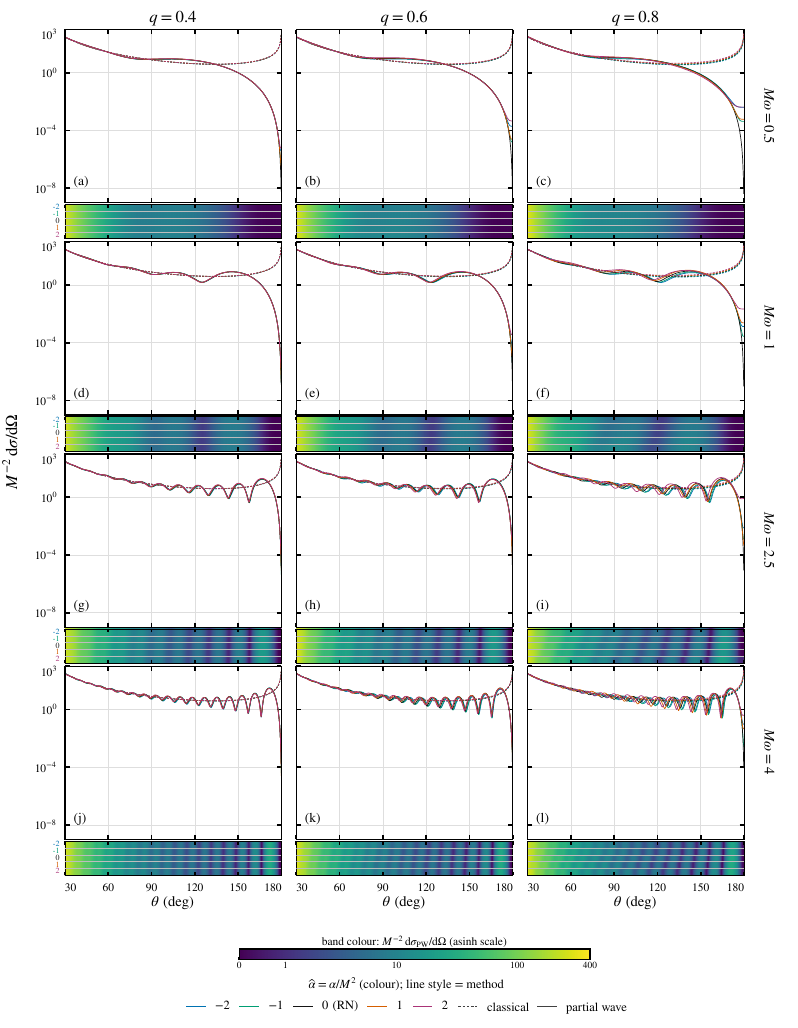}
 \caption{Differential scattering for $q=0.4,0.6,0.8$ (columns) and $\wM=0.5,1,2.5,4$ (rows).
Colours follow Figure~\ref{fig:1}; solid/dashed curves show partial-wave/classical results.
Grey-separated intensity strips run from $\aM=-2$ to $2$ in unit steps,
top to bottom.
Their common asinh scale spans $0\leq I/M^2\leq400$ with characteristic intensity $M^2$.}
 \label{fig:5}
\end{figure}

Figure~\ref{fig:5} shows interference between reflected angular modes over $30^\circ\leq\theta\leq180^\circ$.
Coupling changes their amplitudes and phases,
shifting the fringes rather than simply rescaling the RN curve.
Higher frequencies produce finer angular structure.
The scalar orbiting interpretation~\cite{ref40} provides a semiclassical comparison;
our amplitudes also retain vector polarisation.
We average the classical cross sections of the two optical metrics:
\begin{equation}
 \frac{\dd\sigma_{\rm cl}^{\rm unpol}}{\dd\Omega}
 =\frac12\sum_{p={\rm o,e}}\sum_i
 \frac{b_i^p}{\sin\theta}
 \left|\frac{\dd b_i^p}{\dd\theta}\right|.
 \label{eq:5.1}
\end{equation}
Here $i$ labels distinct ray trajectories reaching angle $\theta$ with impact parameter $b_i^p$ in parity $p$.
This frequency-independent reference omits interference.
It diverges at the backward caustic,
where the wave intensity remains finite.
The solid curves use the coherent amplitudes in Eqs.~\eqref{eq:3.5} and~\eqref{eq:3.6}.
The common strip scale permits absolute-intensity comparisons across charges and frequencies.

We separate the modulus and phase by writing $S_\ell^p=\rho_\ell^p e^{2i\delta_\ell^p}$,
where $\rho_\ell^p=\sqrt{1-\Gamma_\ell^p}$ and $\delta_\ell^p$ is a real reflection phase.
Then
\begin{equation}
 |S_\ell^{\rm o}-S_\ell^{\rm e}|^2
 =(\rho_\ell^{\rm o}-\rho_\ell^{\rm e})^2
 +4\rho_\ell^{\rm o}\rho_\ell^{\rm e}
 \sin^2(\delta_\ell^{\rm o}-\delta_\ell^{\rm e}).
 \label{eq:5.2}
\end{equation}
The two terms measure differences in reflected magnitude and phase.
Even nearly equal absorption probabilities can therefore accompany helicity reversal.
The complex multipole sum in Eq.~\eqref{eq:3.5} also retains their angular interference.

At $\theta=\pi$,
only the reversing contribution survives.
Every nonzero-coupling case studied here has finite backward intensity,
against an exact null for the uncoupled test field.
This distinction is specific to our fixed-background equations.
Coupled electromagnetic-gravitational perturbations can produce RN helicity reversal without the non-minimal interaction~\cite{ref24}.

\begin{figure}[!htbp]
 \centering\includegraphics[width=\textwidth]{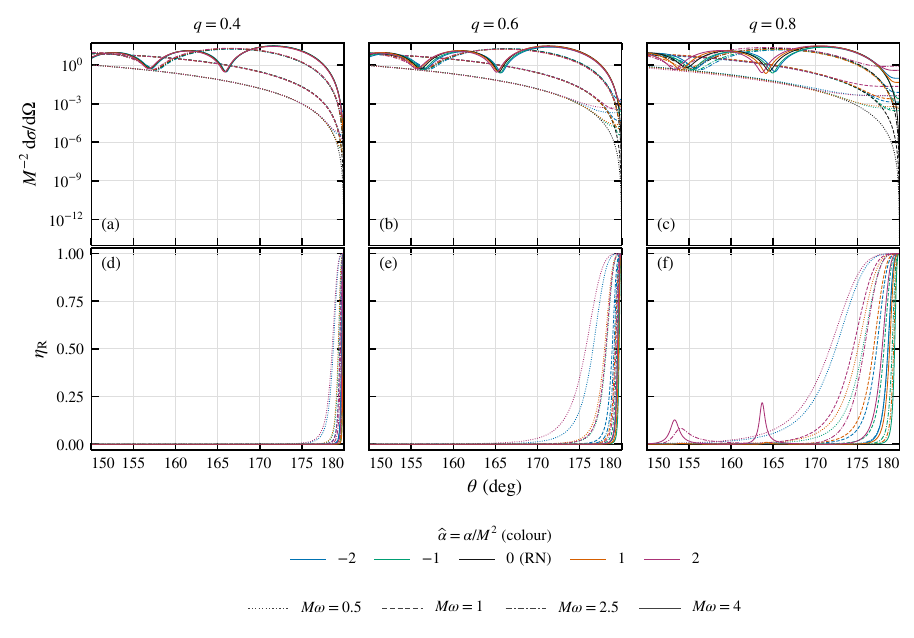}
 \caption{Near-backward intensity (top) and helicity-reversal fraction (bottom) for $q=0.4,0.6,0.8$ (columns).
Colours follow Figure~\ref{fig:1}.
Dotted,
dashed,
dash-dotted and solid curves denote $\wM=0.5,1,2.5,4$.
The RN ratio is undefined at $180^\circ$;
its curve ends at the last defined sample.
Zero intensities are omitted on logarithmic axes.}
 \label{fig:6}
\end{figure}

Figure~\ref{fig:6} compares intensity with the reversed fraction.
The coupled endpoint has $\eta_{\rm R}=1$,
which specifies helicity but not signal strength.
For RN,
both intensities vanish there and the ratio is undefined.

The four frequencies show how rapidly the helicity content changes close to $180^\circ$.
For $q=0.8$,
$\aM=2$ and $\theta=179^\circ$,
we obtain $\eta_{\rm R}=0.9998$,
$0.9990$,
$0.9973$ and $0.9519$ at $\wM=0.5,1,2.5,4$.
At the exact endpoint, the signal is fully reversed for every case.
But the angular region around it has a helicity mix that depends on frequency.

\begin{figure}[!htbp]
 \centering\includegraphics[width=\textwidth]{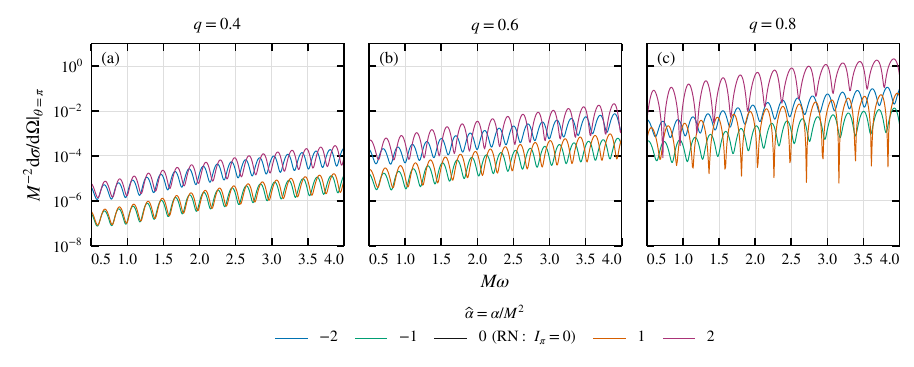}
 \caption{Exact backward intensity $I_\pi\equiv I(\pi)$ versus frequency for $q=0.4,0.6,0.8$ (columns).
Coupling colours follow Figure~\ref{fig:1}.
RN has an exact backward null.}
 \label{fig:7}
\end{figure}

We can see frequency oscillations in the backward spectra of Figure~\ref{fig:7}.
At $q=0.8$ and $\wM=1$,
$I(\pi)/M^2$ is $2.85\times10^{-4}$,
$2.39\times10^{-3}$ and $2.21\times10^{-2}$ for $\aM=-1,1,2$.
At $\wM=4$,
the corresponding values are $2.59\times10^{-3}$,
$4.50\times10^{-2}$ and $3.92\times10^{-1}$.
Positive coupling yields a stronger signal at these frequencies.
The intermediate samples contain maxima and minima,
so a frequency-independent rescaling would miss the phase interference.
For $\aM=-2$,
we obtain $I(\pi)/M^2=1.33\times10^{-3}$ at $\wM=1$ and $9.53\times10^{-2}$ at $\wM=4$.

We also compare charge at fixed coupling and frequency.
At $\aM=1$ and $\wM=1$,
the backward intensities are $1.13\times10^{-7}M^2$,
$1.11\times10^{-5}M^2$ and $2.39\times10^{-3}M^2$ for $q=0.4,0.6,0.8$.
Accordingly, a finite signal at low charge can be far weaker than the high-charge case.
This difference is resolved by the logarithmic axes.

\subsection{Backward-scattering maps and rings}

Rays going around opposite sides of the black hole can end up scattered into almost the same backward direction.
Their wave interference forms rings around that direction.
We use the semiclassical spin-1 glory formula to interpret their spacing~\cite{ref41,ref42},
\begin{equation}
 \left.\frac{\dd\sigma}{\dd\Omega}\right|_{\rm glory}
 \simeq 2\pi\omega b_g^2
 \left|\frac{\dd b}{\dd\Theta}\right|_{\Theta=\pi}
 J_2^2(\omega b_g\sin\theta).
 \label{eq:5.3}
\end{equation}
Here $J_2$ is the Bessel function of the first kind of order two.
We suppress the parity label on the impact parameter $b$ and deflection angle $\Theta$.
For a parity-degenerate field,
this formula gives a helicity-preserving pattern with a dark centre.
The glory impact parameter $b_g$ corresponds to a ray deflected through $\pi$.
Near the backward direction,
the argument of the Bessel function is approximately $\omega b_g(\pi-\theta)$.
At fixed frequency,
a larger $b_g$ therefore places the rings closer to the backward axis and reduces their angular spacing.
Every $J_2$ pattern goes to zero at the centre.
Neither one by itself can give a nonzero backward signal.

The deflection function follows directly from the optical metric:
\begin{equation}
 \Theta_p(b)=2\int_{r_0}^{\infty}
 \frac{b\mathcal W_p(r)\,\dd r}
 {r^2\sqrt{1-b^2H_p(r)}}-\pi,
 \qquad b^2H_p(r_0)=1,
 \label{eq:5.4}
\end{equation}
where $r_0$ is the outer turning point of a scattering ray.
For a fixed scattering angle $\theta$,
the ray contributing to Eq.~\eqref{eq:5.1} satisfy $\Theta_p(b_i^p)=2\pi n+\theta$ or $2\pi(n+1)-\theta$,
where $n=0,1,2,\ldots$ counts increasing winding number.
The primary family of backward rays is selected by solving $\Theta_p(b_g^p)=\pi$.
Its ring spacing is controlled by $\omega b_g^p$,
while $C_p=2\pi(b_g^p)^2|\dd b/\dd\Theta|_{\pi}$ controls the prefactor $\omega C_p$ in Eq.~\eqref{eq:5.3}.
We likewise suppress the parity label on $C$ when the branch is specified.
The derivative measures the range of impact parameters redirected into an angular interval.
The ring scale and brightness coefficient can therefore change by different amounts.
For $q=0$ our ray data give $b_g/M=5.35696$ and $C/M^3=30.7523$,
consistent with the Schwarzschild spin-1 glory scale~\cite{ref25}.

At $(q,\aM)=(0.8,1)$,
we obtain $b_g/M=4.63487,4.62921$ and $C/M^3=30.5061,30.9245$ for odd and even parity.
The even branch has a smaller glory impact parameter but a larger semiclassical coefficient.
For fixed frequency,
any near-backward Bessel feature therefore sits at a slightly bigger angular distance away from the backward axis.
At $(0.8,2)$,
the relative odd-even separation in $b_g$ is $0.679\%$.
while the relative odd-even separation in $C$ is $7.214\%$.
In both cases, we take the average of the two branches as the denominator.
These different responses help explain the changes around the backward direction.
The central reversed intensity is still governed by the difference between the complex scattering amplitudes of the two parities.
Increasing frequency moves a fixed Bessel feature to smaller $\pi-\theta$,
which makes the near-backward rings shrink.

\begin{figure}[!htbp]
 \centering\includegraphics[width=\textwidth]{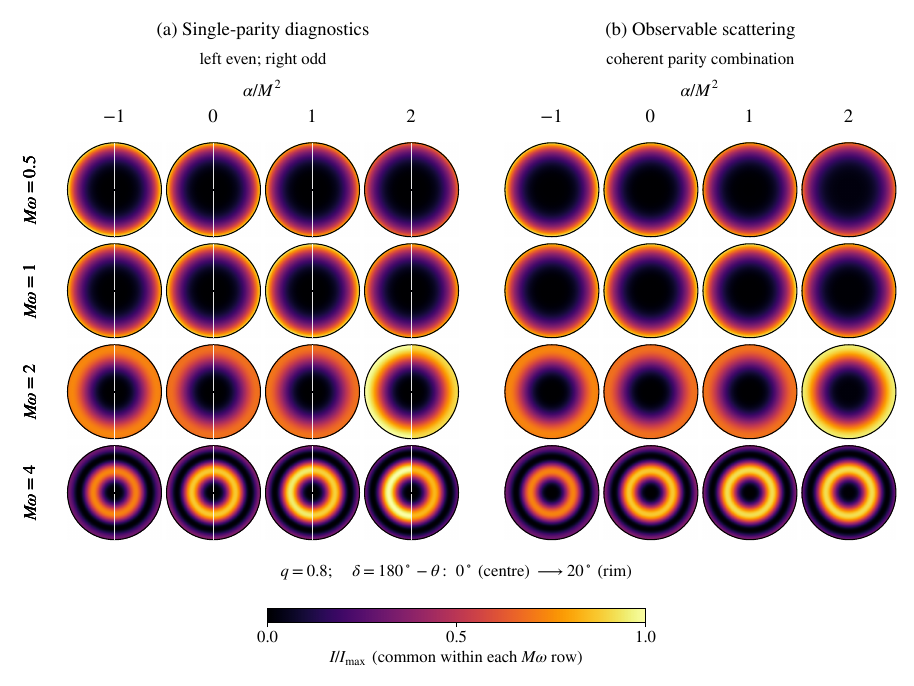}
 \caption{Backward angular maps at $q=0.8$:
columns $\aM=-1,0,1,2$ and rows $\wM=0.5,1,2,4$.
Disc radius spans $0\leq180^\circ-\theta\leq20^\circ$.
(a) Equal-parity diagnostics,
even on each disc's left half and odd on its right.
(b) Physical coherent intensity $I_{\rm P}+I_{\rm R}$.
The linear scale shows $I/I_{\max}$,
with one maximum per frequency row across both arrays and all couplings.}
 \label{fig:8}
\end{figure}

Figure~\ref{fig:8}(a) isolates each parity as a diagnostic by replacing the pair in Eq.~\eqref{eq:3.5} with $(S_\ell^p,S_\ell^p)$.
This removes helicity reversal and gives a dark centre.
Panel (b) uses the actual pair $(S_\ell^{\rm e},S_\ell^{\rm o})$ before taking either modulus squared.
It cannot be recovered by adding the diagnostic intensities.
Odd and even parity label magnetic- and electric-type spherical modes;
a conventional polariser does not isolate either complete set.

The diagnostic and physical patterns coincide for RN.
With coupling,
the physical pattern contains both modified preserving rings and a reversing contribution.
Each frequency row shares one normalisation across both arrays and all couplings.
Figure~\ref{fig:7} resolves weak central intensities quantitatively.

\section{Shadow geometry and split capture rims}\label{sec:6}

\begin{table}[!htbp]
 \centering
 \footnotesize
 \setlength{\tabcolsep}{2.5pt}
 \renewcommand{\arraystretch}{1.05}
 \begin{tabular*}{\textwidth}{@{\extracolsep{\fill}}r*{6}{r}@{}}
 \toprule
 \multicolumn{7}{c}{\textit{Full critical impact parameters}}\\
 \midrule
 & \multicolumn{3}{c}{$b_c^{\rm o,full}/M$} & \multicolumn{3}{c}{$b_c^{\rm e,full}/M$}\\
 \cmidrule(lr){2-4} \cmidrule(lr){5-7}
 $\aM$ & $q=0.4$ & $q=0.6$ & $q=0.8$ & $q=0.4$ & $q=0.6$ & $q=0.8$\\
 \midrule
 $-2$ & 5.098857 & 4.979083 & 4.820661 & 5.098077 & 4.974087 & 4.798789\\
 $-1.5$ & 5.087487 & 4.949802 & 4.757492 & 5.087039 & 4.946820 & 4.743363\\
 $-1$ & 5.076050 & 4.919974 & 4.690796 & 5.075846 & 4.918566 & 4.683503\\
 $-0.5$ & 5.064547 & 4.889602 & 4.620341 & 5.064494 & 4.889227 & 4.618195\\
 $0$ & 5.052977 & 4.858694 & 4.545986 & 5.052977 & 4.858694 & 4.545986\\
 $0.5$ & 5.041343 & 4.827263 & 4.467773 & 5.041289 & 4.826836 & 4.464659\\
 $1$ & 5.029645 & 4.795328 & 4.386010 & 5.029422 & 4.793496 & 4.370528\\
 $1.5$ & 5.017883 & 4.762914 & 4.301344 & 5.017369 & 4.758488 & 4.256578\\
 $2$ & 5.006060 & 4.730053 & 4.214736 & 5.005125 & 4.721580 & 4.105671\\
 \bottomrule
 \end{tabular*}
 \vspace{1.1ex}\par
 \begin{tabular*}{\textwidth}{@{\extracolsep{\fill}}r*{9}{r}@{}}
 \toprule
 \multicolumn{10}{c}{\textit{First- and second-order critical impact parameters}}\\
 \midrule
 & \multicolumn{3}{c}{$b_c^{(1)}/M$} & \multicolumn{3}{c}{$b_c^{{\rm o},(2)}/M$} & \multicolumn{3}{c}{$b_c^{{\rm e},(2)}/M$}\\
 \cmidrule(lr){2-4} \cmidrule(lr){5-7} \cmidrule(lr){8-10}
 $\aM$ & $q=0.4$ & $q=0.6$ & $q=0.8$ & $q=0.4$ & $q=0.6$ & $q=0.8$ & $q=0.4$ & $q=0.6$ & $q=0.8$\\
 \midrule
 $-2$ & 5.099385 & 4.983383 & 4.851227 & 5.098865 & 4.979196 & 4.820169 & 5.098013 & 4.972796 & 4.779178\\
 $-1.5$ & 5.087783 & 4.952211 & 4.774917 & 5.087491 & 4.949855 & 4.757447 & 5.087011 & 4.946256 & 4.734389\\
 $-1$ & 5.076181 & 4.921038 & 4.698607 & 5.076051 & 4.919992 & 4.690842 & 5.075838 & 4.918392 & 4.680594\\
 $-0.5$ & 5.064579 & 4.889866 & 4.622296 & 5.064547 & 4.889605 & 4.620355 & 5.064493 & 4.889205 & 4.617793\\
 $0$ & 5.052977 & 4.858694 & 4.545986 & 5.052977 & 4.858694 & 4.545986 & 5.052977 & 4.858694 & 4.545986\\
 $0.5$ & 5.041375 & 4.827522 & 4.469676 & 5.041343 & 4.827260 & 4.467735 & 5.041290 & 4.826860 & 4.465173\\
 $1$ & 5.029773 & 4.796350 & 4.393366 & 5.029643 & 4.795303 & 4.385602 & 5.029430 & 4.793703 & 4.375354\\
 $1.5$ & 5.018172 & 4.765178 & 4.317056 & 5.017879 & 4.762823 & 4.299586 & 5.017400 & 4.759223 & 4.276529\\
 $2$ & 5.006570 & 4.734006 & 4.240746 & 5.006050 & 4.729819 & 4.209688 & 5.005197 & 4.723419 & 4.168697\\
 \bottomrule
 \end{tabular*}
 \vspace{1.1ex}\par
 \begin{tabular*}{\textwidth}{@{\extracolsep{\fill}}r*{6}{r}@{}}
 \toprule
 \multicolumn{7}{c}{\textit{Glory impact parameters}}\\
 \midrule
 & \multicolumn{3}{c}{$b_g^{\rm o}/M$} & \multicolumn{3}{c}{$b_g^{\rm e}/M$}\\
 \cmidrule(lr){2-4} \cmidrule(lr){5-7}
 $\aM$ & $q=0.4$ & $q=0.6$ & $q=0.8$ & $q=0.4$ & $q=0.6$ & $q=0.8$\\
 \midrule
 $-2$ & 5.259773 & 5.138154 & 4.970656 & 5.259338 & 5.135428 & 4.958962\\
 $-1.5$ & 5.250287 & 5.114157 & 4.919916 & 5.250038 & 5.112557 & 4.912691\\
 $-1$ & 5.240752 & 5.089803 & 4.867225 & 5.240640 & 5.089061 & 4.863680\\
 $-0.5$ & 5.231168 & 5.065089 & 4.812474 & 5.231139 & 5.064895 & 4.811490\\
 $0$ & 5.221534 & 5.040010 & 4.755556 & 5.221534 & 5.040010 & 4.755556\\
 $0.5$ & 5.211852 & 5.014564 & 4.696377 & 5.211822 & 5.014351 & 4.695135\\
 $1$ & 5.202119 & 4.988748 & 4.634867 & 5.201999 & 4.987855 & 4.629209\\
 $1.5$ & 5.192338 & 4.962561 & 4.570990 & 5.192063 & 4.960451 & 4.556327\\
 $2$ & 5.182507 & 4.936002 & 4.504762 & 5.182009 & 4.932058 & 4.474300\\
 \bottomrule
 \end{tabular*}
 \caption{Critical and glory impact parameters in units of $M$ for three charges and nine uniformly spaced couplings.
\textit{Full} uses the complete optical factors;
$(1)$ and $(2)$ denote the coupling expansions in Eq.~\eqref{eq:6.3}.
First-order splitting vanishes;
second-order normalised splitting is $x_0^2/2$.
Values are rounded to six decimal places.}
 \label{tab:1}
\end{table}

Table~\ref{tab:1} compares photon ring and backward scattering radii.
For each charge,
both phonton rings are shrunk by positive coupling and enlarged by negative coupling compared with RN.
The even-parity radius is smaller than the odd-parity radius.
Equation~\eqref{eq:2.8} shows $H_{\rm e}\geq H_{\rm o}$ in the exterior region,
so the maximum of $H_{\rm e}$ is at least as large as that of $H_{\rm o}$.
Equation~\eqref{eq:2.12} then gives $b_c^{\rm e}\leq b_c^{\rm o}$.
Glory radii shift the same way along the table,
but their shifts are different, because $b_g$ is determined by a deflection integral.

We normalise the mean radius shift and edge separation to the same-charge RN radius:
\begin{equation}
 \delta_{\rm sh}=
 \frac{(b_c^{\rm o}+b_c^{\rm e})/2-b_c^{\rm RN}}{b_c^{\rm RN}},
 \qquad
 \Delta_{\rm rim}=\frac{b_c^{\rm o}-b_c^{\rm e}}{b_c^{\rm RN}}.
 \label{eq:6.1}
\end{equation}
At $q=0.8$ and $\aM=1$,
the average shift is $-3.689\%$ and the separation is $0.341\%$.
For $\aM=-1$,
they are $+3.105\%$ and $0.160\%$.
Across the charge grid at $\aM=2$,
the average shift changes from $-0.938\%$ to $-8.486\%$,
while the separation grows from $0.0185\%$ to $2.3991\%$.
Thus the parity contrast grows faster with charge than the common shift (Table~\ref{tab:2}).

We can expand around the RN photon ring to explain this.
The RN photon ring radius is $r_{\rm ph}^{\rm RN}=(3M+\sqrt{9M^2-8Q^2})/2$.
We define $x_0=4\alpha Q^2/(r_{\rm ph}^{\rm RN})^4$.
The function $f/r^2$ has a stationary point at $r_{\rm ph}^{\rm RN}$;
a small perturbation of the orbit radius does not affect this maximum at first order.
Thus the common correction in Eq.~\eqref{eq:2.8} gives
\begin{equation}
 \delta_{\rm sh}=-\frac{x_0}{2}+O(x_0^2),\qquad
 \Delta_{\rm rim}=\frac{x_0^2}{2}+O(x_0^3),
 \label{eq:6.2}
\end{equation}
for fixed subextremal charge and small coupling.
At first order, the two shadow merge at $b_c^{\rm RN}(1-x_0/2)$.
At second order, the radii listed in Table~\ref{tab:1} include the shift of the photon orbit:
\begin{equation}
 \frac{b_c^{{\rm o},(2)}}{b_c^{\rm RN}}=1-\frac{x_0}{2}+B x_0^2,
 \qquad
 \frac{b_c^{{\rm e},(2)}}{b_c^{\rm RN}}=1-\frac{x_0}{2}+\left(B-\frac12\right)x_0^2,
 \label{eq:6.3}
\end{equation}
where $B=3/8-(r_{\rm ph}^{\rm RN}-M)/(2r_{\rm ph}^{\rm RN}-3M)$.
Their difference gives Eq.~\eqref{eq:6.2}.
At $q=0.8$ and $\aM=2$,
the second-order even radius is $4.168697M$,
compared with the full value $4.105671M$.
We therefore use the complete optical factors at finite coupling.
The leading mean shift changes sign with $\alpha$,
but the quadratic splitting does not.
Higher orders account for the unequal finite-coupling separations at opposite signs of $\alpha$.

The geometric absorption and shadow are related by their definitions,
\begin{equation}
 \frac{\sigma_{\rm geo}^{\rm unpol}}{\pi(b_c^{\rm RN})^2}
 =(1+\delta_{\rm sh})^2+\frac{\Delta_{\rm rim}^2}{4}.
 \label{eq:6.4}
\end{equation}
The leading fractional change in geometric absorption is twice the average fractional radius shift,
and this relation applies to the high-frequency region.

\begin{table}[!htbp]
 \centering
 \footnotesize
 \setlength{\tabcolsep}{2.5pt}
 \renewcommand{\arraystretch}{1.05}
 \begin{tabular*}{\textwidth}{@{\extracolsep{\fill}}r*{6}{r}@{}}
 \toprule
 \multicolumn{7}{c}{\textit{Absorption at $M\omega=1$}}\\
 \midrule
 & \multicolumn{3}{c}{$\Delta\sigma_{\rm abs}\,[\%]$} & \multicolumn{3}{c}{$\mathcal A_{\rm abs}\,[\%]$}\\
 \cmidrule(lr){2-4} \cmidrule(lr){5-7}
 $\aM$ & $q=0.4$ & $q=0.6$ & $q=0.8$ & $q=0.4$ & $q=0.6$ & $q=0.8$\\
 \midrule
 $-2$ & $+1.28$ & $+2.33$ & $+11.04$ & $0.0154$ & $0.0608$ & $0.1709$\\
 $-1.5$ & $+0.94$ & $+1.73$ & $+9.40$ & $0.0084$ & $0.0318$ & $0.1170$\\
 $-1$ & $+0.62$ & $+1.15$ & $+7.27$ & $0.0036$ & $0.0133$ & $0.0809$\\
 $-0.5$ & $+0.30$ & $+0.58$ & $+4.28$ & $0.0009$ & $0.0032$ & $0.0377$\\
 $0$ & $0$ & $0$ & $0$ & $0$ & $0$ & $0$\\
 $0.5$ & $-0.29$ & $-0.62$ & $-5.91$ & $0.0008$ & $0.0033$ & $0.1360$\\
 $1$ & $-0.57$ & $-1.32$ & $-13.14$ & $0.0032$ & $0.0151$ & $0.8373$\\
 $1.5$ & $-0.84$ & $-2.11$ & $-20.25$ & $0.0070$ & $0.0412$ & $2.1569$\\
 $2$ & $-1.10$ & $-3.05$ & $-25.20$ & $0.0119$ & $0.0946$ & $2.8813$\\
 \bottomrule
 \end{tabular*}
 \vspace{1.1ex}\par
 \begin{tabular*}{\textwidth}{@{\extracolsep{\fill}}r*{6}{r}@{}}
 \toprule
 \multicolumn{7}{c}{\textit{Shadow displacement and parity splitting}}\\
 \midrule
 & \multicolumn{3}{c}{$\delta_{\rm sh}\,[\%]$} & \multicolumn{3}{c}{$\Delta_{\rm rim}\,[\%]$}\\
 \cmidrule(lr){2-4} \cmidrule(lr){5-7}
 $\aM$ & $q=0.4$ & $q=0.6$ & $q=0.8$ & $q=0.4$ & $q=0.6$ & $q=0.8$\\
 \midrule
 $-2$ & $+0.900$ & $+2.426$ & $+5.802$ & $0.0154$ & $0.1028$ & $0.4811$\\
 $-1.5$ & $+0.679$ & $+1.844$ & $+4.497$ & $0.0089$ & $0.0614$ & $0.3108$\\
 $-1$ & $+0.455$ & $+1.247$ & $+3.105$ & $0.0040$ & $0.0290$ & $0.1604$\\
 $-0.5$ & $+0.228$ & $+0.632$ & $+1.612$ & $0.0010$ & $0.0077$ & $0.0472$\\
 $0$ & $0$ & $0$ & $0$ & $0$ & $0$ & $0$\\
 $0.5$ & $-0.231$ & $-0.651$ & $-1.755$ & $0.0011$ & $0.0088$ & $0.0685$\\
 $1$ & $-0.464$ & $-1.323$ & $-3.689$ & $0.0044$ & $0.0377$ & $0.3406$\\
 $1.5$ & $-0.700$ & $-2.017$ & $-5.874$ & $0.0102$ & $0.0911$ & $0.9847$\\
 $2$ & $-0.938$ & $-2.735$ & $-8.486$ & $0.0185$ & $0.1744$ & $2.3991$\\
 \bottomrule
 \end{tabular*}
 \vspace{1.1ex}\par
 \begin{tabular*}{\textwidth}{@{\extracolsep{\fill}}r*{3}{r}@{}}
 \toprule
 \multicolumn{4}{c}{\textit{Backward scattering at $M\omega=1$}}\\
 \midrule
 & \multicolumn{3}{c}{$I(\pi)/M^2$}\\
 \cmidrule(lr){2-4}
 $\aM$ & $q=0.4$ & $q=0.6$ & $q=0.8$\\
 \midrule
 $-2$ & $2.14\times10^{-6}$ & $4.89\times10^{-5}$ & $1.33\times10^{-3}$\\
 $-1.5$ & $6.56\times10^{-7}$ & $1.64\times10^{-5}$ & $7.51\times10^{-4}$\\
 $-1$ & $1.26\times10^{-7}$ & $3.67\times10^{-6}$ & $2.85\times10^{-4}$\\
 $-0.5$ & $7.63\times10^{-9}$ & $2.79\times10^{-7}$ & $3.48\times10^{-5}$\\
 $0$ & $0$ & $0$ & $0$\\
 $0.5$ & $7.20\times10^{-9}$ & $4.93\times10^{-7}$ & $1.10\times10^{-4}$\\
 $1$ & $1.13\times10^{-7}$ & $1.11\times10^{-5}$ & $2.39\times10^{-3}$\\
 $1.5$ & $5.59\times10^{-7}$ & $7.97\times10^{-5}$ & $1.18\times10^{-2}$\\
 $2$ & $1.74\times10^{-6}$ & $3.60\times10^{-4}$ & $2.21\times10^{-2}$\\
 \bottomrule
 \end{tabular*}
 \caption{Physical signatures on the grid of Table~\ref{tab:1}.
Absorption,
asymmetry and backward intensity use $\wM=1$.
Here $\Delta\sigma_{\rm abs}=\sigma_{\rm abs}^{\rm unpol}/\sigma_{\rm abs}^{\rm RN}-1$ compares equal mass and charge;
Eq.~\eqref{eq:6.1} defines the shadow measures.
The first four observables are percentages;
backward intensities have three significant figures.}
 \label{tab:2}
\end{table}

Table~\ref{tab:2} shows absorption, scattering and shadow results together.
Intermediate couplings illustrate that the finite coupling response is not simply a linear shift.
For $q=0.8$,
increasing $\aM$ from $0.5$ to $1.5$ pushes the absorption reduction at $\wM=1$ from $5.91\%$ up to $20.25\%$,
within this range, the shadow separation grows from $0.0685\%$ to $0.9847\%$.

\begin{figure}[!htbp]
 \centering\includegraphics[width=\textwidth]{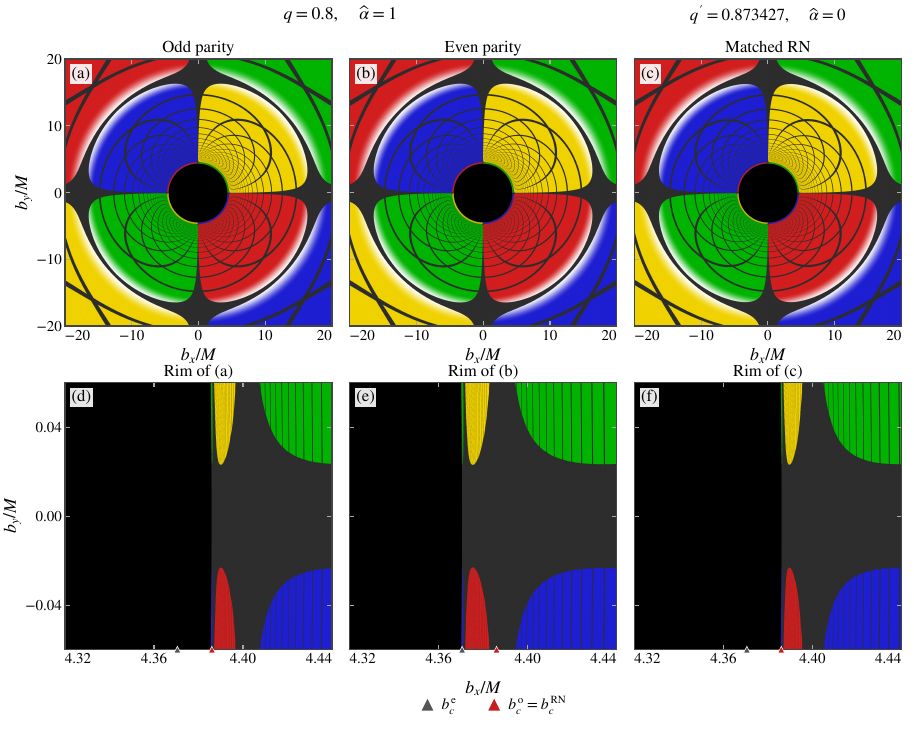}
 \caption{Ray-traced shadow mimicry at $r_{\rm obs}=100M$.
(a,b) Odd/even images at $q=0.8$,
$\aM=1$.
(c) Same-mass RN with $q'\simeq0.873427$ matching $b_c^{\rm o}$.
(d--f) Corresponding edge enlargements on common impact-parameter coordinates $(b_x,b_y)$.
Grey/red triangles mark $b_c^{\rm e}$/$b_c^{\rm o}=b_c^{\rm RN}(q')$.
Black denotes capture;
colours show escape directions on a four-quadrant sphere with a $6^\circ$ mesh and $4^\circ$ axial spot.}
 \label{fig:9}
\end{figure}

Figure~\ref{fig:9} compares the coupled parity images with an RN black hole whose charge matches the odd capture radius.
The coloured background records escape direction~\cite{ref43,ref44}.
Both parities are captured inside $b_c^{\rm e}$;
only odd parity is captured between $b_c^{\rm e}$ and $b_c^{\rm o}$.
On the common enlarged window,
the RN and odd edges coincide while the even edge remains smaller.
The background patterns also differ.

\section{Physical interpretation and limitations}\label{sec:7}

\subsection{Matching shadow size by changing charge}

Changing RN charge can reproduce one coupled shadow radius,
but it cannot reproduce parity splitting.
Nor does radius matching imply equal absorption or angular scattering spectra.

Previous work has used matching optical scales to identify similar wave signatures in different black hole geometries.
De Paula et al. matched the critical impact parameter $b_c$ of Ay\'on-Beato--Garc\'ia and RN black holes to find similar massless-scalar absorption spectra~\cite{ref45}.
They subsequently matched the glory impact parameter $b_g$ to compare their angular scattering patterns~\cite{ref46}.
The spectra can be close without being identical.
Their later study of massive scalar waves showed how the field mass affects such similarities between regular and RN black holes~\cite{ref47}.
Tang et al. also compared absorption and scattering with matched critical or glory impact parameters for Frolov and other black hole backgrounds~\cite{ref48}.
Here we test a propagation-induced change against the uncoupled RN and use parity differences to distinguish cases with the same shadow radius.
This follows the distinction between modified propagation and modified geometry studied in curvature-coupled black hole imaging~\cite{ref10}.

We choose $q'$ at fixed mass so that $b_c^{\rm RN}(q')=b_c^{\rm o}(q,\aM)$.
The unmatched even capture area then differs by
\begin{equation}
 \delta_{\rm e}^{\rm geo}
 =\frac{\pi[b_c^{\rm RN}(q')]^2-\pi(b_c^{\rm e})^2}
 {\pi(b_c^{\rm e})^2}
 =\left(\frac{b_c^{\rm o}}{b_c^{\rm e}}\right)^2-1.
 \label{eq:7.1}
\end{equation}
Table~\ref{tab:3} uses the grid of Tables~\ref{tab:1} and~\ref{tab:2}.
Every target admits a subextremal RN match,
with $q'<q$ for negative coupling and $q'>q$ for positive coupling.
At $q=0.8$,
the $\aM=-2$ and $2$ cases require $q'=0.629945$ and $0.937395$,
leaving residuals $0.9136\%$ and $5.3835\%$.
For Figure~\ref{fig:9},
$q'=0.873427$ leaves a $0.7097\%$ area residual.
Its denominator is the even capture area;
the $0.341\%$ edge separation instead uses the original same-charge RN radius.

\begin{table}[!htbp]
 \centering
 \footnotesize
 \setlength{\tabcolsep}{2.5pt}
 \renewcommand{\arraystretch}{1.05}
 \begin{tabular*}{\textwidth}{@{\extracolsep{\fill}}r*{6}{r}@{}}
 \toprule
 \multicolumn{7}{c}{\textit{RN matching charge and even-channel residual}}\\
 \midrule
 & \multicolumn{3}{c}{$q'$} & \multicolumn{3}{c}{$\delta_{\rm e}^{\rm geo}\,[\%]$}\\
 \cmidrule(lr){2-4} \cmidrule(lr){5-7}
 $\aM$ & $q=0.4$ & $q=0.6$ & $q=0.8$ & $q=0.4$ & $q=0.6$ & $q=0.8$\\
 \midrule
 $-2$ & $0.331497$ & $0.488261$ & $0.629945$ & $0.0306$ & $0.2010$ & $0.9136$\\
 $-1.5$ & $0.349872$ & $0.518338$ & $0.675510$ & $0.0176$ & $0.1206$ & $0.5966$\\
 $-1$ & $0.367337$ & $0.546858$ & $0.718902$ & $0.0080$ & $0.0573$ & $0.3117$\\
 $-0.5$ & $0.384015$ & $0.574027$ & $0.760367$ & $0.0021$ & $0.0153$ & $0.0929$\\
 $0$ & $0.400000$ & $0.600000$ & $0.800000$ & $0$ & $0$ & $0$\\
 $0.5$ & $0.415370$ & $0.624894$ & $0.837748$ & $0.0022$ & $0.0177$ & $0.1395$\\
 $1$ & $0.430190$ & $0.648798$ & $0.873427$ & $0.0089$ & $0.0765$ & $0.7097$\\
 $1.5$ & $0.444510$ & $0.671782$ & $0.906751$ & $0.0205$ & $0.1861$ & $2.1144$\\
 $2$ & $0.458376$ & $0.693897$ & $0.937395$ & $0.0374$ & $0.3592$ & $5.3835$\\
 \bottomrule
 \end{tabular*}
 \vspace{1.1ex}\par
 \begin{tabular*}{\textwidth}{@{\extracolsep{\fill}}r*{6}{r}@{}}
 \toprule
 \multicolumn{7}{c}{\textit{Photon-orbit ratios}}\\
 \midrule
 & \multicolumn{3}{c}{$\eta_{\rm o}$} & \multicolumn{3}{c}{$\eta_{\rm e}$}\\
 \cmidrule(lr){2-4} \cmidrule(lr){5-7}
 $\aM$ & $q=0.4$ & $q=0.6$ & $q=0.8$ & $q=0.4$ & $q=0.6$ & $q=0.8$\\
 \midrule
 $-2$ & $0.996788$ & $0.999187$ & $1.027822$ & $0.995564$ & $0.990642$ & $0.985432$\\
 $-1.5$ & $0.992660$ & $0.986588$ & $0.993608$ & $0.991960$ & $0.981537$ & $0.966313$\\
 $-1$ & $0.988591$ & $0.974254$ & $0.958614$ & $0.988275$ & $0.971895$ & $0.944652$\\
 $-0.5$ & $0.984585$ & $0.962271$ & $0.923739$ & $0.984505$ & $0.961651$ & $0.919704$\\
 $0$ & $0.980646$ & $0.950727$ & $0.890338$ & $0.980646$ & $0.950727$ & $0.890338$\\
 $0.5$ & $0.976777$ & $0.939715$ & $0.860156$ & $0.976694$ & $0.939031$ & $0.854713$\\
 $1$ & $0.972983$ & $0.929322$ & $0.834995$ & $0.972645$ & $0.926448$ & $0.809525$\\
 $1.5$ & $0.969267$ & $0.919632$ & $0.816084$ & $0.968494$ & $0.912839$ & $0.747832$\\
 $2$ & $0.965633$ & $0.910714$ & $0.803505$ & $0.964236$ & $0.898026$ & $0.650109$\\
 \bottomrule
 \end{tabular*}
 \caption{Charge matching on the grid of Table~\ref{tab:1}.
At fixed mass,
$q'$ satisfies $b_c^{\rm RN}(q')=b_c^{\rm o}(q,\aM)$.
All matches are subextremal.
Equation~\eqref{eq:7.1} defines the percentage even-area residual;
Eq.~\eqref{eq:4.2} defines the orbital ratios.
Matching capture radii does not assume equal wave spectra.}
 \label{tab:3}
\end{table}

Matching capture area need not match the orbital escape ratio.
For the RN spacetime,
$\eta_{\rm RN}=\sqrt{1-2Q'^2/(r_{\rm ph}^{\rm RN})^2}$ lies between $1/\sqrt2$ and $1$.
Here $Q'=Mq'$ is the comparison charge,
and $r_{\rm ph}^{\rm RN}$ is computed for that charge.
The smallest value is reached at extremality.
The even branch at $(q,\aM)=(0.8,2)$ gives $\eta_{\rm e}=0.6501$.
Its leading coefficient $8\pi\eta_{\rm e}e^{-\pi\eta_{\rm e}}=2.119$ is larger than the RN maximum of $1.927$.
No RN charge can match both scales for this parity.
At $(q,\aM)=(0.8,-2)$,
the odd parity ratio $\eta_{\rm o}=1.0278$ is larger than the RN upper limit of $1$.
Even at $(q,\aM)=(0.8,1)$,
the matched RN ratio $0.850860$ differs from the coupled values $0.834995$ and $0.809525$.
Spectral oscillations therefore provide information beyond shadow size.

At fixed nonzero charge and regular frequency,
the $O(\alpha^2)$ potential difference gives $S_\ell^{\rm o}-S_\ell^{\rm e}=O(\alpha^2)$ for analytic scattering.
The reversing amplitude starts at this order,
but its intensity starts at $O(\alpha^4)$.
Our finite-coupling results use the full equations.
Unlike the Weyl interaction~\cite{ref09},
the Einstein-tensor correction vanishes as $Q\to0$ at finite $\alpha$.

The effective optical description connects our results with birefringence in nonlinear electrodynamics~\cite{ref14,ref15}.
There, the optical metrics are derived from perturbations of a nonlinear electromagnetic theory.
In the Euler--Heisenberg example,
the nonlinear interaction also changes the black hole background~\cite{ref15}.
By contrast, our model couples a distinct vector test field to curvature on a fixed RN background.
This common optical description does not imply equivalence of the underlying theories.

At fixed finite $\aM$,
small charge suppresses the common correction as $\aM q^2$ and parity splitting as $\aM^2q^4$.
Our parameter grid characterizes the propagation response rather than astrophysical charge distributions.
Any application therefore requires an estimate of $Q$,
together with a statement of whether the charge is electromagnetic or arises from a different sector.

\subsection{Relating the results to observations}

Our images describe test-field capture rather than electromagnetic brightness.
An observational comparison requires a mechanism that produces and detects the field,
its source polarisation and absolute flux,
and an instrument response.
Image modelling also requires mass,
distance and an emission model.
A large reversed fraction near a dark minimum alone does not establish detectability.

Photon subrings have characteristic long-baseline signatures~\cite{ref49}.
To apply these imaging techniques to our field, one must adopt the source and detector models outlined above.
The calculated shifts are not observational exclusion limits for $\alpha$.

We retain only the interaction in Eq.~\eqref{eq:2.1} on fixed RN spacetime.
Regular radial equations alone do not justify neglecting other interactions.
Higher-derivative Einstein-Maxwell studies show that consistency depends on the operators and physical setup~\cite{ref50}.
Our larger-coupling examples characterise this specified model over finite parameter and frequency ranges.
A microscopic theory is needed to establish its validity,
and dedicated source and instrument modelling is needed for observational inference.

\section{Conclusions}\label{sec:8}

We demonstrate charge-activated birefringence for a neutral vector field coupled to the Einstein tensor on RN spacetime.
Direct integration of the radial equations shows that positive coupling raises both barriers and suppresses low-frequency and transition-region transmission.
Negative coupling reverses this trend,
and the response strengthens with charge.

The common response dominates the parity contrast at moderate coupling.
For $q=0.8$ and $\aM=1$,
unpolarised absorption falls by $13.14\%$ at $\wM=1$ and the average shadow radius falls by $3.69\%$.
The absorption asymmetry and edge separation are only $0.837\%$ and $0.341\%$.
Even so,
the difference between complex parity amplitudes produces a fully helicity-reversed backward signal,
where the uncoupled test-field intensity vanishes.

Changing RN charge can mimic one shadow radius but not both parity edges.
For the same example,
$q'\simeq0.873427$ matches the odd radius while leaving a $0.7097\%$ even-area residual.
The orbital escape ratios differ as well
Measurements of shadow size together with parity-sensitive absorption, backward scattering and spectral oscillations thus break the degeneracy for this model.
To connect these signatures to observations, one still needs production and detection mechanisms for the field, source modelling and instrument response.

\begin{acknowledgments}
This work was supported by the Zhejiang normal university Doctorial research fund Contract No. ZC302924005.
\end{acknowledgments}

\section*{Data availability}
The data are available from the authors upon reasonable request.

\bibliographystyle{JHEP}
\bibliography{refs}
\end{document}